# Pressure sensor-based tongue-placed electrotactile biofeedback for balance improvement - Biomedical application to prevent pressure sores formation and falls

N. Vuillerme, O. Chenu, N. Pinsault, A. Moreau-Gaudry, A. Fleury, J. Demongeot and Y. Payan

*Abstract*— We introduce the innovative technologies, based on the concept of "sensory substitution", we are developing in the fields of biomedical engineering and human disability. Precisely, our goal is to design, develop and validate practical assistive biomedical and/or technical devices and/or rehabilitating procedures for persons with disabilities, using artificial tongue-placed tactile biofeedback systems.

Proposed applications are dealing with: (1) pressure sores prevention in case of spinal cord injuries (persons with paraplegia, or tetraplegia); and (2) balance control improvement to prevent fall in older and/or disabled adults.

This paper describes the architecture and the functioning principle of these biofeedback systems and presents preliminary results of two feasibility studies performed on young healthy adults.

## I. INTRODUCTION

THE present paper introduces the innovative technologies, based on the concept of "sensory substitution" [1], we are developing in the fields of biomedical engineering and human disability. Precisely, our goal is to design, develop and validate practical assistive biomedical and/or technical devices and/or rehabilitating procedures for persons with disabilities, using artificial tongue-placed electrotactile biofeedback systems.

Proposed applications are dealing with: (1) pressure sores prevention in case of spinal cord injuries (persons with paraplegia, or tetraplegia); and (2) balance control improvement to prevent fall in older and/or disabled adults.

The tongue-placed electrotactile output device ("Tongue Display Unit" - TDU), on which we focused our attention, was initially introduced by Bach-y-Rita and colleagues [2] and used as a tactile-vision sensory substitution system to provide distal spatial information to blind individuals [3]. It consists in a 2D array of miniature electrodes (12 × 12 matrix) held between the lips and positioned in close contact with the anterior-superior surface of the tongue. A flexible cable connects the matrix to an external electronic device delivering the electrical signals that individually activate the electrodes and therefore the tactile receptors of the tongue (Figure 1A).

At this point, however, to provide a perspective for the application of this device outside the laboratory framework and to permit its use over long-time period in real-life environment, this device had to be ergonomically and esthetically acceptable. The current ribbon TDU system did not meet these requirements.

Within this context, with the help of the companies Coronis-Systems and Guglielmi Technologies Dentaires, we have developed a wireless radio-controlled version of the 6 × 6 TDU matrix. This consists in a matrix glued onto the inferior part of the orthodontic retainer including microelectronics, antenna and battery, which can be worn inside the mouth like a dental retainer (Figure 1B). Note that it was decided to reduce by a factor of 4 the overall size of the Bach-y-Rita's device since the addressed biomedical applications do not require the dense 12 × 12 TDU resolution of the tactile visual substitution systems.

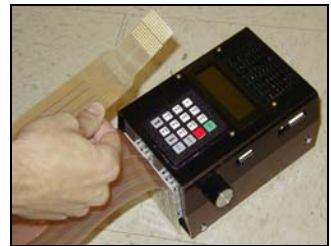 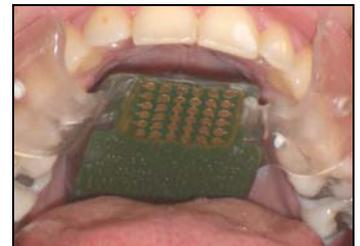

**Fig. 1A.** Wire TDU: 12 × 12 matrix and flexible cable [2]

**Fig. 1B.** Wireless TDU: 6 × 6 matrix and embedded radio-frequency circuit

The following sections describe the architecture and the functioning principle of our biofeedback systems for:

(1) pressure sores prevention (Section II) and

(2) balance improvement for fall prevention (Section III), and presents preliminary results of two feasibility studies performed on young healthy adults.

Manuscript received April 2, 2007. This work was supported in part by the company IDS, Floralis (Université Joseph Fourier, Grenoble), and the Fondation Garches. The company Vista Medical is acknowledged for supplying the FSA pressure mapping systems.

Nicolas Vuillerme is with the TIMC-IMAG Laboratory, UMR UJF CNRS 5525, Faculté de Médecine de Grenoble, Bâtiment Jean Roget, F38706 La Tronche Cédex; phone: +33 4 76 63 74 86; fax: +33 4 76 51 86 67; mail: Nicolas.Vuillerme@imag.fr

Olivier Chenu, Nicolas Pinsault, Alexande Moreau-Gaudry, Anthony Fleury, Jacques Demongeot and Yohan Payan are also with the TIMC-IMAG Laboratory, UMR UJF CNRS 5525, La Tronche, France. Mail: firstname.lastname@imag.fr

## II. PRESSURE SORES PREVENTION

### A. Objective

A pressure sore is defined as an area of localized damage to the skin and underlying tissues caused by overpressure, shearing, friction or a combination of these factors [4]. Located near bony prominences such as the ischium, sacrum and trochanter, pressure sores present a prevalence from 23% to 39% in adults with spinal cord injuries [5,6] and are recognized as the main cause of rehospitalization for patients with paraplegia [7].

Contrary to healthy individuals, i.e. with intact sensory capacities, individuals with spinal cord injuries (persons with paraplegia, or tetraplegia) do not get the feedback arising from buttock area informing them of a localized excess of pressure at the skin/seat interface and the necessity to make adaptive postural correction to prevent pressure sores [8].

Within this context, we developed an original system for preventing the formation of pressure sores in individuals with spinal cord injuries [9]. Its underlying principle consists in supplying through the wireless 6 × 6 TDU the user with supplementary sensory information regarding the adequate seated posture to adopt to prevent excessive local pressures.

The purpose of the following experiment was to assess the performance of this system in young healthy adults.

### B. Experimental procedure

Six young healthy males adults (age = 27.2 ± 3.7 years; body weight = 78.2 ± 9.7 kg; height = 181.7 ± 9.1 cm) volunteered for this experiment and gave their informed consent to the experimental procedure as required by the Helsinki declaration (1964) and the local Ethics Committee. None of the participants presented any history of motor problem, neurological disease or vestibular impairment.

Subjects were seated comfortably in a chair onto which was put a pressure mapping system (FSA Seat 32/63, Vista Medical Ltd.), allowing real-time acquisition of the pressure applied on the seat/skin interface. In a preliminary calibration stage, for each subject, we determined the patterns of the pressure distribution (1) in a normal seated posture and (2) in 8 other different postures, corresponding to the situations in which the subject moves his chest in the 8 front, rear, left, right, front-left, front-right, rear-left, rear-right directions. The experiment was next initiated. For a period randomly ranged from 15 to 35 seconds, pressures applied to the buttock area of the seated subjects were recorded and overpressures zones were localized (Figure 2, upper right panel). Thanks to the preliminary calibration stage, we determined the zone, reachable with a simple movement of the chest forward, backward, on the left or right sides, which would decrease the most these measured overpressures. When this zone was localized, electrical stimulation was provided in 4 distinct zones located in the front, rear, left and right of the matrix, according to the following scheme. If the localized overpressure was supposed to disappear with a postural change in the front, back, left or right direction, the corresponding 4 electrodes of the back, front; right or left row of the matrix were respectively activated (Figure 2, lower left panel).

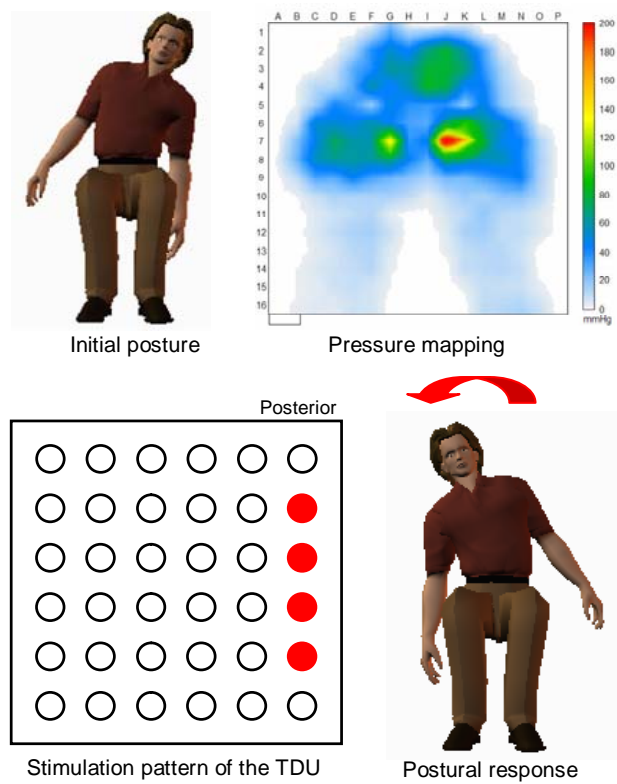

**Fig. 2.** Principle of the biofeedback system for pressure sores prevention.

After each movement following the electro-stimulated direction, a new record of the pressure map was provided. This procedure was repeated 40 times during two experimental sessions separated by a rest period of 5 minutes. Subjects were not given feedback about their performance.

Several practice runs were performed prior to the test to ensure that subjects had mastered the relationship between the lingual stimulations and the postural response to execute.

The pressure map applied at the seat/skin interface was recorded at a frequency of 5 Hz. By computing the differences between the pressure map recorded before and after electro-stimulation, we determined whether or not the postural responses were adapted to the electrotactile stimulation and decreased the pressure distribution applied on the seat/skin interface.

### C. Results

The percentage of adapted postural responses was 95.5%. This result suggested that young healthy adults were able to take advantage of an artificial tongue-placed tactile biofeedback to produce adapted postural behavior to reduce localized excess of pressure at the skin/seat interface in a seated posture.

## III. BALANCE IMPROVEMENT FOR FALL PREVENTION

### A. Objective

Postural control is a particularly complex task involving various sensory and motor components. Among the sensory inputs relevant to the regulation of postural sway, the importance of somatosensory information from the foot sole is now well established [10]. Clinically, alteration or loss of somatosensory information from the lower limbs resulting from normal aging or disease (e.g., diabetic peripheral neuropathy) is known to impair postural control [11]. Progressive degeneration of sensory inputs from the lower extremities also represents a common clinical finding associated with aging [11,12] and has even been identified as important contributing factor to the occurrence of falls in elderly [13,14]. Therefore, it is legitimate to propose that a therapeutic intervention and/or a technical assistance designed to increase somatosensory function of the plantar sole could be of great interest for controlling balance and preventing falls in the elderly and/or disabled persons.

Within this context, we developed an original biofeedback system for improving postural control [15,16]. Its underlying principle consists in supplying the user with supplementary sensory information related to foot sole pressure distribution through the TDU. The purpose of the following experiment was to assess the performance of this system in young healthy adults.

### B. Experimental procedure

The 6 young healthy males adults who participated to the first experiment volunteered for this experiment.

Subjects stood barefoot, foot together, their hands hanging at the sides, with their eyes closed. They were asked to sway as little as possible in two No-TDU and TDU experimental conditions. The No-TDU condition served as a control condition. In the TDU condition, subjects performed the postural task using a plantar pressure-based, tongue-placed tactile biofeedback system. A plantar pressure data acquisition system (FSA Orthotest Mat, Vista Medical Ltd.) was used. The pressure sensors transduced the magnitude of pressure exerted on each left and right foot sole at each sensor location into the calculation of the positions of the resultant ground reaction force exerted on each left and right foot, referred to as the left and right foot centre of foot pressure, respectively ($CoP_{lf}$ and $CoP_{rf}$). The positions of the resultant CoP were then computed from the left and right foot CoP trajectories through the following relation:

$CoP = CoP_{lf} \times R_{lf} / (R_{lf} + R_{rf}) + CoP_{rf} \times R_{rf} / (R_{rf} + R_{lf})$,

where $R_{lf}$, $R_{rf}$, $CoP_{lf}$, $CoP_{rf}$ are the vertical reaction forces under the left and the right feet, the positions of the CoP of the left and the right feet, respectively (red circle, Figure 3, upper right panel).

CoP data were then fed back to the TDU at a 3 Hz frequency.

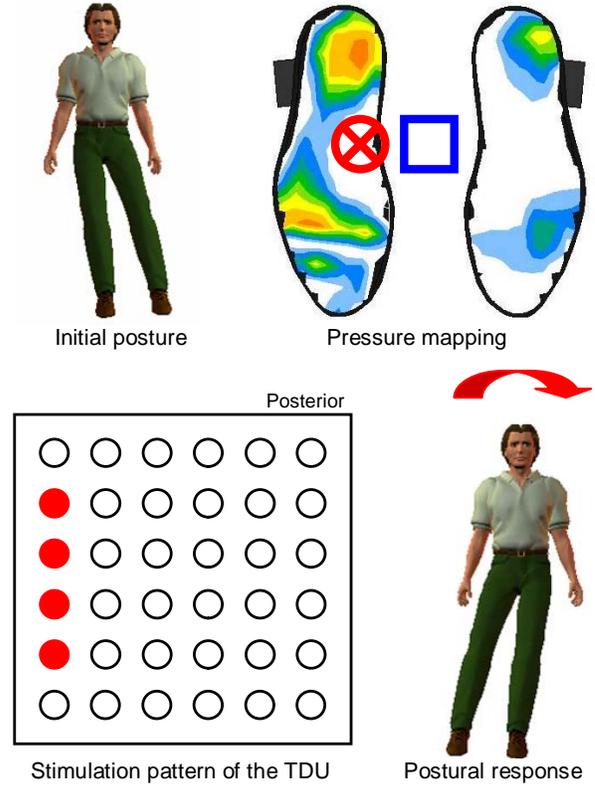

**Fig. 3.** Principle of the biofeedback system for balance improvement and fall prevention.

The underlying principle of our biofeedback system was to supply the users with supplementary information about the position of the CoP relative to a predetermined adjustable "dead zone" (DZ) through the TDU (blue square, Figure 3, upper right panel). In the present experiment, antero-posterior and medio-lateral bounds of the DZ were set as the standard deviation of subject's CoP displacements recorded for 10 s preceding each experimental trial. A simple and intuitive coding scheme for the TDU, consisting in a "threshold-alarm" type of feedback rather that a continuous feedback about ongoing position of the CoP, was then used:

(1) when the position of the CoP was determined to be within the DZ, no electrical stimulation was provided in any of the electrodes of the matrix;

(2) when the position of the CoP was determined to be outside the DZ, electrical stimulation was provided in distinct zones of the matrix, depending on the position of the CoP relative to the DZ. Specifically, four different zones located in the front, rear, left and right of the matrix were defined; the activated zone of the matrix corresponded to the position of the CoP relative to the DZ. For instance, in the case that the CoP was located at the right hand side of the DZ, a stimulation of four electrode located in the right portion of the matrix was provided (red circles, Figure 3, lower left panel).

Several practice runs were performed prior to the test to ensure that subjects had mastered the relationship between the position of the CoP relative to the DZ and lingual stimulations. Three 30 seconds trials for each experimental condition were performed. The order of presentation of the two experimental conditions was randomized. Subjects were not given feedback about their performance.

The means of the three trials performed in each of the two experimental conditions were used for statistical analyses. A one-way ANOVA 2 Conditions (No-TDU vs. TDU) was applied to the CoP surface area data. Level of significance was set at 0.05.

*C. Results*

Analysis of the surface area covered by the trajectory of the CoP showed a main effect of Condition, yielding a narrower surface area in the TDU than No-TDU condition ($F(1,5) = 18.93$, $P < 0.01$).

These results suggested that young healthy adults were able to take advantage of an artificial tongue-placed tactile biofeedback to improve their postural control during quiet standing.

## IV. DISCUSSION AND CONCLUSION

This paper described the architecture and the functioning principle of two pressure sensor-based tongue-placed electrotactile biofeedback systems for (1) pressure sores formation prevention in case of spinal cord injuries (persons with paraplegia, or tetraplegia); and (2) balance improvement for fall prevention in older and/or disabled adults. Preliminary results of two feasibility studies performed on young healthy adults also were presented.

Overall, the present findings evidence that electrotactile stimulation of the tongue can be used (1) as a part of a device designed to prevent pressure sores (experiment 1) and (2) to improve postural control during quiet standing (experiment 2).

Although these feasibility studies have been conducted in young healthy individuals, i.e., in individuals with intact sensory, motor, cognitive capacities, we strongly believe that our results could have significant implications in rehabilitative areas, for enhancing/restoring/preserving balance and mobility in individuals with reduced capacities (resulting either from normal aging, trauma or disease) with the aim at ensuring autonomy and safety in occupations of daily living and maximizing quality of life. Along these lines, the effectiveness of our biofeedback systems in the two biomedical applications presented in this paper is currently being evaluated in paraplegic patients, individuals with somatosensory loss in the feet from diabetic peripheral neuropathy and persons with lower limb amputation, respectively.

ACKNOWLEDGMENT

The authors are indebted to Professor Paul Bach-y-Rita for introducing us to the Tongue Display Unit and for discussions about sensory substitution. Paul has been for us more than a partner or a supervisor: he was a master inspiring numerous new fields of research in many domains of neurosciences, biomedical engineering and physical rehabilitation. The authors would like to thank subject volunteers. Special thanks also are extended to Paul Pavan for technical support and L. Abitite and Claire H. for various contributions.